\documentclass[12pt,a4paper]{article}
\pdfoutput=1
\usepackage{jheppub}
\usepackage{empheq}
\usepackage{amsfonts}
\usepackage{youngtab}

\newcommand{\de}{\partial} 
\newcommand{\ket}[1]{\lvert #1 \rangle} 
\newcommand{\bra}[1]{\langle #1 \rvert} 

\newcommand{\R}{\widetilde{R}}

\newcommand{\be}{\begin{equation}}
\newcommand{\ee}{\end{equation}}
\newcommand{\bea}{\begin{eqnarray}}
\newcommand{\eea}{\end{eqnarray}}

\newcommand{\1}{\scalebox{0.7}{$d$-1}}

 \newcommand{\5}{\scalebox{0.7}{\phantom{1}}}
 \newcommand{\6}{\scalebox{0.7}{1}}
 \newcommand{\7}{\scalebox{0.7}{$d$}}

\def\eqa{&=&} 
\def\ccr{\nonumber\\}
\def\ra{\rangle}
\def\la{\langle}
\mathchardef\mhyphen="2D

\title{Massive and massless higher spinning particles in odd dimensions}
\author[a]{Fiorenzo Bastianelli,}
\author[a]{Roberto Bonezzi,}
\author[b,c]{Olindo Corradini,}
\author[d]{Emanuele Latini}

\affiliation[a] {Dipartimento di Fisica ed Astronomia, Universit{\`a} di Bologna and\\
INFN, Sezione di Bologna, via Irnerio 46, I-40126 Bologna, Italy}
\affiliation[b] {Facultad de Ciencias en F\'isica y Matem\'aticas\\
Universidad Aut\'onoma de Chiapas, Ciudad Universitaria, Tuxtla Guti\'errez 29050, M\'exico}
\affiliation[c]  {Dipartimento di Scienze Fisiche, Informatiche e Matematiche\\ Universit\`a di Modena e Reggio Emilia, Via Campi 213/A, I-41125 Modena, Italy}
\affiliation[d] {Institut f{\"u}r Mathematik, Universit{\"a}t Z{\"u}rich-Irchel, Winterthurerstrasse 190, CH-8057 Z{\"u}rich, Switzerland}

\emailAdd{bastianelli@bo.infn.it}\emailAdd{bonezzi@bo.infn.it}\emailAdd{olindo.corradini@unach.mx}
\emailAdd{emanuele.latini@math.uzh.ch}

\abstract{We study actions for massive bosonic particles of higher spins by dimensionally reducing an action for
massless particles.  For the latter we take a model with a SO($N$) extended local supersymmetry
on the worldline, that is known to describe massless (conformal) particles of higher spins in flat spacetimes of
even dimensions. Dimensional reduction produces an action for massive spinning particles in odd dimensions.
The  field equations that emerge in a quantization \`a la Dirac are shown to be equivalent to the Fierz-Pauli ones.
The massless limit generates a multiplet of massless states with higher spins, whose
first quantized field equations have a geometric form with fields belonging to various types of Young tableaux.
These geometric equations can be partially integrated to show their equivalence with the standard
Fronsdal-Labastida equations.
We covariantize our model to check whether an extension to curved spacetimes can be achieved.
Restricting to (A)dS spaces, we find that the worldline gauge algebra becomes nonlinear,
but remains first class. This  guarantees consistency on such backgrounds.
A light cone analysis confirms the presence of the expected propagating degrees of freedom.
A covariant analysis is worked out explicitly for the massive case, which is seen to give rise
to the Fierz-Pauli equations extended to (A)dS spaces. It is worth noting that in $D=3$ the massless limit
of our model with $N\to \infty$  has the same field content of the Vasiliev's theory
that accommodates each spin exactly once.}


\begin{document}
\maketitle
\flushbottom


\section{Introduction}

Higher spin field theories have recently been the focus of much interest. One of the main motivations arises from the study
of the interacting higher spin field equations found by Vasiliev~\cite{Vasiliev:1990en,Vasiliev:1995dn,Vasiliev:2003ev}. They
involve an infinite number of higher spin fields on AdS spaces, and
find interesting applications in AdS/CFT dualities  \cite{Sezgin:2002rt,Klebanov:2002ja, Sezgin:2003pt, Giombi:2009wh, Gaberdiel:2010pz,Didenko:2012tv, Giombi:2013fka}.
For an introduction to these subjects see, for example, the reviews
 ~\cite{Bekaert:2005vh, Bekaert:2010hw, Sagnotti:2011qp, Didenko:2014dwa, Vasiliev:2014vwa}  and references therein.

One approach to study free higher spin fields in flat and curved backgrounds is to analyze the first quantization of relativistic particles. This approach was followed in  \cite{Bastianelli:2007pv, Bastianelli:2008nm, Bastianelli:2012bn},  which focused on
massless spinning particles in even spacetime dimensions.
In those references a certain class of higher spin states, defined by the  SO($N$) spinning particle action of  \cite{Gershun:1979fb, Howe:1988ft, Howe:1989vn},
was analyzed in great details. The covariant quantization of the spinning particles was analyzed in flat and (A)dS spaces to show
how well-known higher spin (HS) field equations would emerge from the Dirac quantization  procedure.
In addition, the path integral quantization was used to give a worldline representation of  the
one-loop effective action
on (A)dS spaces, allowing for the calculation of the heat kernel coefficients corresponding to the divergencies of the effective action in $D=4$.

The standard SO($N$) spinning particle action describes massless (in fact, conformal \cite{Siegel:1988ru, Siegel:1988gd})
particles of higher spin in spacetimes of even dimensions. They can be coupled to  conformally flat backgrounds
\cite{Bastianelli:2008nm}, which include in particular
(A)dS spaces \cite{Kuzenko:1995mg}. However, in odd dimensions the model is empty.
The  gauging of the full SO($N$) extended worldline supersymmetry, which is the defining property of the model, constrains
the propagating degrees of freedom to be that of a pure massless particle of
spin $s=\frac{N}{2}$, but forces at the same time the spacetime to be even dimensional (this happens for $N>2$ \emph{i.e.} $s>1$).
One could gauge a subgroup of the SO($N$) symmetry group to describe a multiplet of spinning particles, a fact which
is even desirable within the prospect of introducing interactions, but that would prevent an extension to curved backgrounds,
as the gauging  of the SO($N$) charges is instrumental in providing a first class algebra on curved spaces \cite{Bastianelli:2008nm}.

Here we wish to continue the analysis of HS fields within the worldline approach
and set ourselves to study bosonic massive and massless higher spinning particles in odd dimensions.
We consider the introduction of a mass by dimensional reduction of the SO($N$) spinning particle,
and provide some solutions to the problems mentioned previously.
By construction, the emerging model has massive degrees of freedom in odd dimensions only.
Taking the massless limit gives a multiplet of HS particles in odd dimensions.
The analysis of the physical degrees of freedom  carried by the particle is performed both through a light cone approach and through a covariant approach.
The latter is accomplished by using the Dirac quantization method. It shows how the dynamics is described in a gauge invariant way through
linearized curvatures (which may be expressed in terms of gauge potentials if desired).
Having performed the analysis in flat space,  we proceed by noting
that the gauge algebra can be covariantized to include (A)dS backgrounds, while keeping it first class. This
provides a consistent model for both the massive and massless cases on (A)dS  spaces.
We perform again a light cone analysis to confirm that the particle propagates the same degrees of freedom
as in flat space. A covariant analysis is carried out explicitly for the massive case, and we find that the covariant Dirac
constraints  can be reduced to the Fierz-Pauli equations extended to (A)dS spaces.  A similar treatment
of the massless case is more complex, and we content ourself to report the explicit example of $s=2$ in $D=3$.
It is interesting to note that in the massless case and for integer  $s=\frac{N}{2}\to \infty$  one finds the same field content
appearing in the three dimensional Vasiliev's's theory, where each integer spin occurs precisely once.
The coupling to (A)dS is presumably crucial for having a chance of
studying the interactions carried by the Vasiliev's  model in a first quantized approach.

\label{sec:intro}

\section{Review of the SO(N) massless spinning particles}\label{review}
We start reviewing the action of the SO($N$) massless spinning particle, that describes  a particle of
spin $s=\frac{N}{2}$ upon quantization.
This is mainly to introduce notations, conventions and methods.
 We consider the motion on a flat $(D+1)$-dimensional Minkowski
space $M_{D+1}$ and take $N$ even to restrict ourselves
to bosonic particles of integer spin $s$. The dynamical variables of the spinning particle are given by:
\begin{itemize}
\item the cartesian coordinates  $x^m$  of the particle on ${M}_{D+1}$ (${m=0,1,..,D}$)
\item  their conjugate momenta $p_m$
\item $N$ real Grassmann variables
with spacetime vector indices $\psi_i^m $ ($i =1,..,N$)
\item
the SO($N$)-extended supergravity multiplet on the worldline, whose gauge fields are made up  by
the einbein $e$, the $N$ gravitinos $\chi_i$, and the SO($N$) gauge field  $a_{ij}$.
\end{itemize}
The phase space action of the model, $S = \int d\tau L$, is identified by the  lagrangian\footnote{The  Minkowski metric
$\eta_{mn}\sim (-,+,\cdots,+)$  is  used to raise and lower spacetime indices.
Indices named $m,n,..$ etc.  refer to spacetime indices ($m,n=0,1,...D$), while those named $i,j,..$ etc. stand for
internal SO($N$) indices ($i,j=1,.., N$).}
\be
L =
p_m \dot x^m +\frac{i}{2}  \psi_{im} \dot \psi_i^m
-e \underbrace{\Big ( \frac12 p_m p^m \Big )}_{\cal H}
- i \chi_i \underbrace{\Big ( p_m \psi^m_i\Big )}_{{\cal Q}_i}
- \frac12 a_{ij} \underbrace{\Big (i \psi_i^m \psi_{jm} \Big )}_{{\cal J}_{ij}}
\label{action1}
\ee
where ${\cal H}, {\cal Q}_i, {\cal J}_{ij}$ denote first class constraints gauged by  $e,\chi_i,a_{ij}$.
The kinetic term defines the phase space symplectic structure and fixes the graded Poisson brackets
\be
\{x^m, p_n \}_{_{PB}}=\delta^m_n \;, \qquad
\{\psi_i^m,\psi_j^n\}_{_{PB}} = -i \eta^{mn} \delta_{ij}
\ee
(other independent brackets vanish). With them one computes the constraint algebra
\be
 \addtolength{\fboxsep}{5pt}
\begin{gathered}
\{{\cal Q}_i,{\cal Q}_j\}_{_{PB}} = -2i \delta_{ij} {\cal H} \\[1.5mm]
\{{\cal J}_{ij},{\cal Q}_k\}_{_{PB}} = \delta_{jk} {\cal Q}_i -\delta_{ik} {\cal Q}_j  \\[1.5mm]
 \{{\cal J}_{ij},{\cal J}_{kl}\}_{_{PB}} = \delta_{jk} {\cal J}_{il} - \delta_{ik} {\cal J}_{jl}
- \delta_{jl} {\cal J}_{ik} + \delta_{il} {\cal J}_{jk}
\end{gathered}
\label{linear-constraint-algebra}
\ee
which is first class. This algebra is known as the SO($N$)-extended susy
algebra in one dimension,  as it contains $N$ real susy charges  ${\cal Q}_i$. They transform in the vector
representation of SO($N$) (the so-called R-symmetry group) generated by the ${\cal J}_{ij}$ charges,
and close on the Hamiltonian $\cal H$.

In a quantization \`a la Dirac, the constraint functions  $C^A := ( {\cal H},  {\cal Q}_i,  {\cal J}_{ij})$ become operators
that produce the massless higher spin (HS) field equations.
One may write them as $C^A |R\ra=0$, where $|R\ra$ is a vector of the extended Hilbert space.
The solutions of these constraint equations  make up the subspace of physical states.
The remaining Schr\"odinger equation implies that the physical fields do not depend on the time parameter  $\tau$.
These constraint equations have the property of being conformally invariant \cite{Siegel:1988ru, Siegel:1988gd}, and take the form of
the  Bargmann-Wigner equations \cite{Bargmann:1948ck},
studied in arbitrary spacetime dimensions in \cite{Bekaert:2002dt, Bekaert:2006ix}.
Let us describe them. The physical states $|R\ra$  are  contained in a  tensor
\be
R_{m^1_1..m^1_d, ..., m^s_1..m^s_d}
\label{a}
\ee
with $s$ blocks of $d=\frac{D+1}{2}$ antisymmetric indices\footnote{We separate different blocks of antisymmetric indices
by commas;  $D+1$ must be even for nontrivial solutions so that  $d=\frac{D+1}{2}$ is integer.} that
satisfies the properties:\\
$(i)$ it is completely traceless and has the symmetries of a Young tableau with $d$ rows and $s$ columns
(this follows from the $\cal J$ constraints)
\begin{center}
\vspace{10mm}
\begin{picture}(0,60)(30,-35)
\put(0,10){$\yng(4,4,4)$}
\put(0,8){$\underbrace{\hspace*{56pt}}_{s}$}
\put(-115,27){$R_{m^1_1..m^1_d, ..., m^s_1..m^s_d}\sim \scalebox{0.7}{$d$}\,\left\{\rule{0pt}{25pt}\right.$}
\end{picture}
\end{center}\vspace{-30mm}
\begin{equation}\label{yt-tensor-R}
\end{equation}
\\[3mm]
$(ii)$ it satisfies integrability conditions (from half of the $\cal Q$ constraints)
\be
\partial_{[m} R_{m^1_1..m^1_d], ..., m^s_1..m^s_d} =0 \ ,
\label{b}
\ee
interpreted as Bianchi identities once solved,\\
$(iii)$ it satisfies Maxwell equations (from the other half of the $\cal Q$ constraints)
\be
\partial^{m} R_{m m^1_2..m^1_d, ..., m^s_1..m^s_d} =0  \ .
\label{c}
\ee
The $\cal H$ constraint is automatically satisfied as consequence of the constraint algebra.
These are geometric equations for free conformal fields of integer spin $s$,
equivalent to the massless Bargmann-Wigner equations.
They are called geometric as the tensors $R$ can be interpreted as (linearized) curvatures, as we are going to show
later on.

To derive these equations  it is useful to take complex combinations of
the $N=2s$ indices and define (for $I=i=1,..,s$)
\be
\psi_I = \frac{1}{\sqrt{2}}(\psi_i + i \psi_{i+s}) \;, \qquad
\bar \psi^I =\frac{1}{\sqrt{2}}(\psi_i - i \psi_{i+s}) \;.
\ee
Their  non trivial quantum anticommutators  are given by
\be
\{\psi_I^m,\bar \psi^{Jn} \} = \eta^{mn} \delta_I^J
\ee
and describe a set of fermionic creation/annihilation operators.
In this basis only the subgroup U($s$) $\subset$ SO($2s$) is manifest.
The susy charges take the  form
${\cal Q}_I= \psi_I^m p_m$ and $\bar {\cal Q}^I= \bar \psi^{I m} p_m$,
and the susy algebra breaks up into
\be
\addtolength{\fboxsep}{5pt}
\boxed{
\begin{gathered}
\{{\cal Q}_I,\bar {\cal Q}^J\} = 2 \delta_I^J {\cal H} \ ,
\qquad \{{\cal Q}_I,{\cal Q}_J\}  = \{\bar {\cal Q}^I,\bar {\cal Q}^J\} =0 \ .
\end{gathered}
}
\ee
The SO($N$) generators split as ${\cal J}_{ij}\sim ({\cal J}_{I \bar J},{\cal J}_{IJ},  {\cal J}_{\bar I \bar J})
:= ({\cal J}_I{}^J, {\cal K}_{IJ}, \bar {\cal K}^{IJ})$,  which we normalize as
\be
{\cal J}_I{}^J =\psi_I \cdot \bar \psi^J -d\, \delta_I^J
\ , \quad
{\cal K}_{IJ} = \psi_I\cdot \psi_J
\ , \quad
\bar {\cal K}^{IJ} = \bar \psi^I \cdot \bar \psi^J
\ee
(note that  ${\cal J}_I{}^J $ for fixed $I=J$ is a hermitian operator with real eigenvalues).
Then, the SO($N$) algebra breaks up into
\be
\addtolength{\fboxsep}{5pt}
\boxed{
\begin{gathered}
[{\cal J}_I{}^J, {\cal J}_K{}^L] =
\delta_K^J {\cal J}_I{}^L -\delta_I^L {\cal  J}_K{}^J
\\[1.5mm]
[{\cal J}_I{}^J, {\cal K}_{KL}] =
 \delta_K^J {\cal K}_{IL} + \delta_L^J {\cal K}_{KI}
\\[1.5mm]
[{\cal J}_I{}^J,\bar {\cal K}^{KL}] =
 - \delta_I^K \bar {\cal K}^{JL} - \delta_I^L \bar {\cal K}^{KJ}
\\[1.5mm]
[{\cal K}_{IJ},\bar {\cal K}^{KL}] =  \delta_J^K {\cal J}_I{}^L - \delta_J^L {\cal  J}_I{}^K
-\delta_I^K{\cal J}_J{}^L + \delta_I^L {\cal  J}_J{}^K
\end{gathered}
}
\ee
with other commutators vanishing.
The first line identifies the manifest U($s$) subalgebra.
Finally, the remaining non trivial part of the constraint algebra takes the form
\be
\addtolength{\fboxsep}{5pt}
\boxed{
\begin{gathered}
[{\cal J}_I{}^J,{\cal Q}_K] =
\delta_K^J {\cal Q}_I
\\[1.5mm]
[{\cal J}_I{}^J,\bar {\cal Q}^K] =
- \delta_I^K \bar {\cal Q}^J
\\[1.5mm]
[ \bar {\cal K}^{IJ},{\cal Q}_K]=
 \delta_K^J \bar {\cal Q}^I  - \delta_K^I \bar {\cal Q}^J
\\[1.5mm]
[{\cal K}_{IJ},\bar {\cal Q}^K] =
 \delta_J^K {\cal Q}_I  - \delta_I^K {\cal Q}_J  \ .
\end{gathered}
}
\ee

We now analyze the constraints in a quantization \`a la Dirac.
The fermionic operators can be treated using a basis of fermionic coherent states,
so that they can be realized  by letting
$\psi_I^m$ act as multiplication by the Grassmann variables  $\psi_I^m$, and
$\bar \psi^I_m$ as derivation by the Grassmann variable $\psi_I^m$
(\emph{i.e.} $\bar \psi^I_m = \frac{\partial\ }{\partial \psi_I^m}$;
we refrain from denoting operators with a hat, as no confusion can arise).
Using in addition the coordinate representation for the position and momentum operators, one may describe
a generic state $| R\ra$ of the full Hilbert space by the wave function
\be
R(x, \psi) =
(\la x| \otimes \la \psi|) | R\ra =
\sum_{A_i=0}^{D+1}
R_{m_1..m_{A_1}, ...,\, n_1..n_{A_s}}(x)\,
\psi_1^{m_1}..\psi_1^{m_{A_1}}... \psi_s^{n_1}..\psi_s^{n_{A_s}}
\ee
which contains all possible tensors with $s$ blocks of indices, completely antisymmetric in each block.

In the chosen representation the SO($N$) generators take the form
\be
{\cal J}_I{}^J =\psi_I \cdot \frac{\partial}{\partial \psi_J} -d\, \delta_I^J
\ , \quad
{\cal K}_{IJ} = \psi_I\cdot \psi_J
\ , \quad
\bar {\cal K}^{IJ} = \frac{\partial}{\partial \psi_I} \cdot
\frac{\partial}{\partial \psi_J} \;.
\ee
The operator ${\cal J}_I{}^I$ at fixed $I$ counts the number of fermions $\psi^m_I$ of flavor $I$
minus $d$ (this constant arises from a graded-symmetric quantum ordering prescription),
while ${\cal J}_I{}^J$ removes from the wavefunction a fermion $\psi^m_J$ and replaces it with a fermion
$\psi^m_I$ (the fermions of each species antisymmetrize the corresponding indices of the tensor that
multiplies them). In addition,
 ${\cal K}_{IJ} = \psi_I\cdot \psi_J = \psi_I^m\, \eta_{mn} \, \psi^n_J$ acts on the various tensors by adding one index
in the  $I$-th block and one index the $J$-th block by multiplying with the metric tensor $\eta_{mn}$,
each block being then automatically antisymmetrized.
Similarly,
$\bar {\cal K}^{IJ} = \frac{\partial}{\partial \psi_I} \cdot \frac{\partial}{\partial \psi_J}=
\frac{\partial}{\partial \psi^m_I} \, \eta^{mn}\,  \frac{\partial}{\partial \psi^n_J}$
computes traces by contracting one index of the
$I$-th block with one index the $J$-th block through the metric tensor.
Then, it is easy to see that the corresponding constraints  imply
\begin{empheq}[box=\fbox]{align}
{\cal J}_I{}^I |R\ra=0  \  ( I\   {\rm fixed})
\quad
&\Rightarrow \quad
R = R_{m_1..m_d, ...,\, n_1..n_d}(x)\,
\psi_1^{m_1}..\psi_1^{m_d}... \psi_s^{n_1}..\psi_s^{n_d}
\label{j-diag}
\\[1.5mm]
{\cal J}_I{}^J |R\ra=0   \ (I\neq J)
\quad
&\Rightarrow \quad
R\ {\rm satisfies\ algebraic\ Bianchi\ identities}
\label{j-off}
\\[1.5mm]
\bar {\cal K}^{I J} |R\ra =0
\quad
&\Rightarrow \quad
R\ {\rm traceless}
\label{k-bar}
\\[1.5mm]
{\cal K}_{IJ}  |R\ra =0
\quad
&\Rightarrow \quad
R\ {\rm traceless} \;.
\label{k}
 \end{empheq}
Similarly, the constraints $Q_i=({\cal Q}_I, \bar Q^I)$ produce
\begin{empheq}[box=\fbox]{align}
{\cal Q}_I |R\ra =0
\quad &\Rightarrow \quad
 R\ {\rm closed\ (integrability\ conditions\ \to \ potentials)}
\label{q}
\\[1.5mm]
\bar {\cal Q}^I |R\ra =0
\quad &\Rightarrow \quad
R\ {\rm co\mhyphen closed\
(Maxwell\ equations)} \;.
\label{bar-q}
  \end{empheq}
The constraint $\cal H$ is automatically satisfied as a consequence
of the algebra.

Note that the constraints (\ref{j-diag}) and  (\ref{j-off}) correspond to the
generators of the manifest U($s$) $\subset$ SO($2s$).
The tensor $R$ solving these equations has $s$ blocks with $d$ antisymmetric indices each,
consequence of (\ref{j-diag}),
and satisfies algebraic Bianchi identities of the form
\be
R_{[m_1..m_d,n_1]..n_d,...} =0
\ee
where $[...]$ indicates antisymmetrization, consequence of (\ref{j-off}).
There is also a symmetry under an exchange of the blocks. It
can be proved by using finite SO($s$) $\subset$ U($s$) rotations.
For example, a $\frac{\pi}{2}$ rotation in the $I\mhyphen J$ plane,  that implements  $\psi_I \to \psi_J$ and
$\psi_J \to -\psi_I$, implies symmetry under the exchange of block $I$ with block $J$.
Note that the fermionic Fock vacuum $|\Omega\ra  \sim  \Omega(x)$
is not invariant under  [U(1)]$ ^s$ $\subset$ U($s$), as all generators
${\cal J}_I{}^I$ with fixed $I$ transform it by an infinitesimal phase
(${\cal J}_I{}^I |\Omega\ra  = - d|\Omega\ra$).
It is the vector $|R\ra$ in eq. (\ref{j-diag}) that is invariant.
Summarizing, the constraints ${\cal J}_I{}^J$, \emph{i.e.} those belonging to  U($s$), select an irreducible representation of the
general linear group GL($D+1$) identified by a Young tableau
with $d=\frac{D+1}{2}$ rows and $s=\frac{N}{2}$ columns, as depicted in (\ref{yt-tensor-R}).

The constraint $\bar {\cal K}^{IJ}$ removes all possible traces from this tensor, and produces an irreducible representation
of the Lorentz group SO($D,1$). The constraints due to ${\cal K}_{IJ}$ do not give  new independent relations: they say
that pieces equivalent to pure traces must vanish.
The equivalence of  ${\cal K}_{IJ}$ and $\bar {\cal K}^{IJ}$ constraints
is not a consequence of the algebra, but  can be viewed
as a consequence of a duality symmetry enjoyed by the spinning particle.
Indeed, one can  realize the Hodge operator acting in the $I$-th block by
\be
\star_I : \psi_I \leftrightarrow \bar \psi^I \ , \qquad (\star_I)^2 = 1   \ .
\ee
The exchange $ \psi_I \leftrightarrow \bar \psi^I$ maps
the lowest state (in the fermionic Fock vacuum) with the highest state, and so on,
and it is seen to correspond to a dualization of the antisymmetric indices of the tensor $R$ belonging to the $I$-th block.
It is obtained by a discrete O($N$) symmetry transformation
(that reflects one real $\psi_i$ fermion).
Denote now  ${\star}_{IJ}= {\star}_I {\star}_J $ (this combined
transformation can be done within SO($N$)).
Then
\be
{\cal K}_{IJ}|R\ra =0
\quad \Rightarrow \quad
({\star}_{IJ}\,
{\cal K}_{IJ}\,
{\star}_{IJ})\,
( {\star}_{IJ}\, |R\ra)
 = \bar {\cal K}^{I J}  |R^{(\star_{IJ})} \ra  = 0 \ ,
\ee
which implies that $R^{(\star_{I J})}$
is traceless when contracting an index of the $I$-th block with an
index of the $J$-th block.  By $R^{(\star_{I J})}$ we indicate the tensor
dual to $R$ in both set of indices, those of the block $I$ and those of the block $J$.
 Using $\epsilon \epsilon \sim \delta ... \delta$,  one may check that
 tracelessness of $R^{(\star_{I J})}$  implies tracelessness  of $R$ as well.
Finally, note that the $\bar {\cal Q}^I$ constraint is a consequence of
(\ref{q}) and (\ref{k-bar}), since the $ [\bar {\cal K}^{IJ},{\cal Q}_K]=
\delta_K^J \bar {\cal Q}^I - \delta_K^I \bar {\cal Q}^J$.

We have verified that an independent set of constraints is given by (${\cal J}_I{}^J,{\cal Q}_I, \bar{\cal K}^{IJ}$).
They can be implemented in that order to make contact with the Fonsdal-Labastida
formulation of higher spin fields (with or without compensators)
for the particular spin representations carried by the SO($N$) particle.
Let us review these last steps as well.
Gauge potentials $|\phi\ra$ can be introduced by integrating the ${\cal Q}_I$ constraint as
\be
|R\ra =  q |\phi\ra
\label{int-1}
\ee
where $q = {\cal Q}_1 {\cal Q}_2..{\cal Q}_s$. This follows from the nilpotency of the ${\cal Q}_I$'s
together with a Poincar\'e lemma stating that the related cohomologies are trivial in Minkowski space
(all closed forms are exact). Then, the constraints ${\cal J}_I{}^J$ are implemented by taking $|\phi\ra$
to satisfy
\be
{\cal J}_I{}^J |\phi\ra =-  \delta_I{}^J|\phi\ra
\label{int-2}
\ee
that fixes $|\phi\ra$ to contain an irreducible tensor  under GL($D+1$) with Young tableau of the form
\begin{center}
\vspace{10mm}
\begin{picture}(0,60)(30,-35)
\put(0,10){$\yng(4,4)$}
\put(0,8){$\underbrace{\hspace*{56pt}}_{s}$}
\put(-65,19){$\phi\sim \scalebox{0.7}{$(d-1)$}\,\left\{\rule{0pt}{15pt}\right.$}
\end{picture}
\end{center}
\vspace{-10mm}
Finally, the remaining constraints  $\bar {\cal K}^{IJ}$ (the trace constraints)
implement the dynamical equations. One computes
\bea
\bar {\cal K}^{IJ}\, |R\ra =
\bar {\cal K}^{IJ}\, q |\phi\ra =
q^{IJ} \underbrace{
\Big [ -2 {\cal H} + {\cal Q}_I  \bar {\cal Q}^{I} + \frac{1}{2} {\cal Q}_I {\cal Q}_J \bar {\cal K}^{IJ} \Big]}_{ {\cal G}}
|\phi\ra =0
\eea
where $q^{IJ} := \frac{\partial}{{\cal Q}_I} \frac{\partial}{{\cal Q}_J} q$ and $\cal G$ is
the Fronsdal-Labastida operator\footnote{It corresponds to the
Fronsdal kinetic operator for higher spin fields in $D=4$ \cite{Fronsdal:1978rb},
extended to higher dimensions for generic tensors of the Lorentz group by Labastida \cite{Labastida:1987kw}.}
 which is manifestly U($s$) invariant
(one checks that $[{\cal J}_I{}^J,{\cal G}]=0$).
The product of $s+1$ ${\cal Q}_I$'s must vanish, so that one may partially integrate this last equation
to obtain the  Fronsdal-Labastida equation with compensators
\be
\addtolength{\fboxsep}{5pt}
\boxed{
\begin{gathered}
{\cal G} |\phi\ra =  {\cal Q}_I {\cal Q}_J {\cal Q}_K  |\rho^{IJK}\ra
\end{gathered}
} \label{FLwith}
\ee
where the right hand side parametrizes an element of the kernel of $q^{IJ}$, and
the compensator  $|\rho^{IJK}\ra$ has a Young tableau of $GL(D+1)$  of the form
\begin{center}
\vspace{10mm}
\begin{picture}(0,60)(30,-35)
\put(0,10){$\yng(5,2)$}
\put(0,8){$\underbrace{\hspace*{28pt}}_{s-3}$}
\put(-80,20){$\rho^{IJK}\sim \scalebox{0.7}{$(d-1)$}\,\left\{\rule{0pt}{15pt}\right.$}
\end{picture}
\end{center}
\vspace{-10mm}
The gauge symmetries of the Fronsdal-Labastida equation with compensators
are given by
\be
\delta |\phi\ra = {\cal Q}_I |\xi^I\ra \;, \qquad
\delta |\rho^{IJK}\ra = \frac{1}{2} \bar {\cal K}^{[IJ} |\xi^{K]}\ra \;.
\ee
A partial gauge fixing can be used to set the compensators to vanish, and one is left with
the original Fronsdal-Labastida  equation
\be
{\cal G} |\phi\ra = 0
\label{FL-without}
\ee
with gauge symmetries generated by traceless gauge parameters.
The use of compensators in this context was discussed in \cite{Francia:2002pt,Sagnotti:2003qa,Bandos:2005mb}.

\section{Dimensional reduction, massive particles, and massless limit}

Massive spinning particles can be obtained by the Scherk-Schwarz mechanism \cite{Scherk:1979zr}
of dimensionally reducing the massless model on a flat spacetime of the form $M_D\times S^1$.
In practice, one gauges the compact direction $x^D$, corresponding to $S^1$,
by imposing the first class constraint $p_D-m=0$.
Setting $x^m=(x^\mu, x^D)$,  $p_m=(p_\mu, p_D)$,  and $\psi_i^m =(\psi_i^\mu, \theta_i)$ one obtains
in flat, odd $D$ dimensions
 \bea
L \eqa
p_\mu \dot x^\mu + \frac{i}{2} \psi_{i\mu} \dot \psi_i^\mu + \frac{i}{2} \theta_{i} \dot \theta_i  \ccr
&& -e \underbrace{\frac12 (p_\mu p^\mu+ m^2)}_{\cal H}
- i \chi_i \underbrace{\Big ( p_\mu \psi^\mu_i + m \theta_i\Big )}_{{\cal Q}_i}
- \frac12 a_{ij} \underbrace{\Big (i \psi_i^\mu \psi_{j\mu} + i \theta_i \theta_j
\Big )}_{{\cal J}_{ij}} \;.
\label{action2}
\eea
 The constraints satisfy again the same algebra written in (\ref{linear-constraint-algebra}), where  SO($N$) is manifest.
 As shown before, this algebra can be
 equivalently written as in  (\ref{j-diag})-(\ref{k}) and (\ref{q})-(\ref{bar-q}), where only the group U($s$) is manifest.
 The latter form is useful to analyze and solve  the quantum constraints.

  \subsection{Light cone analysis}

 Before discussing the covariant treatment of the constraints at the quantum level,
 let us present a light cone analysis to calculate and check the number of
 propagating  physical degrees of freedom.

We define light cone coordinates by $x^\mu=(x^+,x^-,x^a)$ with
$x^\pm= (x^{D-1}\pm x^0)/\sqrt{2}$ and $x^a$ the transverse directions,
so that $ds^2=2 dx^+dx^- + dx^a dx^a$.
Note that vectors have light cone indices such that $p^+=p_-$ and  $p^-=p_+$.

One can set $x^+=\tau$ as gauge fixing condition, dual to the mass shell constraint  ${\cal H}=0$.
The gauge is well-fixed, as $\{x^+, p^2 +m^2\}_{_{PB}}=2 p_-\neq 0$ (recall that
$p_-=p^+$  is assumed to be invertible in light cone analysis). The constraint $ p^2+m^2=0$ is then solved in terms of
$p_+ = -\frac{1}{2p_-} (p^2_{_T} +m^2)$, where $p^2_{_T}= p^a p^a$ is the transverse momentum squared.
The conjugate variables $(x^+, p_+)$ of the phase space are thus eliminated.
The parameter $x^+$ is taken as the time parameter, and
$-p_+ = \frac{1}{2p_-} (p^2_{_T} +m^2)$ is the corresponding hamiltonian.

Then, one can gauge fix the Majorana fermions $\psi^+_i=0$.
The local susy transformations act on the Majorana fermions as
$\delta \psi^m_i=  \{ \psi^m_i, \epsilon^j {\cal Q}_j \}_{_{PB}}= i \epsilon_i p^m$,
so that using the infinitesimal transformations
$\delta \psi^+_i= i \epsilon_i p^+$ (which are non vanishing)
one can set $\psi^+_i=0$.
The gauge is well-fixed, and indeed $\{ \psi^+_i, {\cal Q}_j \} =  -i p^+\delta_{ij} \neq 0$.
One may solve the constraints ${\cal Q}_i =  0$ by setting
$\psi^-_i = -\frac{1}{p_-} ( p_a \psi_i^a + m \theta_i)$,
and the conjugated variables $(\psi^+_i, \psi^-_i)$ are eliminated as independent phase space coordinates.
The coordinates of the reduced phase space are now given by
$(x^-, p_-)$, $(x^a, p^a)$, and $\psi^a_i$, with lagrangian
\bea
L \eqa  p_- \dot x^- + p_a \dot x^a  +\frac{i}{2}  \psi_{ia} \dot \psi_i^a +\frac{i}{2}  \theta_{i} \dot \theta_i
-\frac{1}{2p_-} (p^2_{_T} +m^2) \ccr
&&- \frac12 a_{ij} \underbrace{\Big (i \psi_i^a \psi_{ja} +i \theta_i\theta_j \Big )}_{{\cal J}_{ij}}\;.
\eea
One may try to reduce the phase space further, implementing the last constraint ${\cal J}_{ij}$.
However, this can be done in a simpler way  \`a la Dirac, since it produces purely algebraic constraints.
This implementation proceeds as described previously, when discussing the massless case in $D+1$ even dimensions.
Taking into account the unique quantum ordering of the quantum constraints ${\cal J}_{ij}$,
one finds a sum of irreps of the SO($D-2$) group that fill an irrep of the SO($D-1$)
rotation group corresponding precisely to the polarizations of a massive spin $s$
in $D$ dimensions (in higher dimensions by spin $s$ we mean a multiplet
corresponding to a rectangular Young tableau with $s$ columns and $\frac{D-1}{2}$ rows).
The corresponding degrees of freedom  are counted by using
a ``factor over hook" type of formula, and their number is given by
\be
Dof(D,s) = \frac{Y_t}{Y_h}
\ee
where
\be
Y_t = \prod_{i=1}^{d-1} \frac{(s+i-2)!  (2s+2i -2)!}{(2i-2)!  (2s +i -2)!} \;, \qquad
Y_h = \prod_{i=1}^{d-1} \frac{(s+i-1)!}{(i-1)!} \;, \qquad d:=\frac{D+1}{2} \;.
\ee
In particular, in $D=3$ one finds two degrees of freedom for any spin $s>0$.
Of course, this is identical to the polarizations of a massless spin $s$ particle in one dimension higher, a fact that is rather evident
from the dimensional reduction process.

\subsection{Covariant analysis}\label{covariantsection}

We are now ready to give a covariant  analysis, implementing the constraints of the Dirac quantization scheme.
We can partially solve them to make contact with known relativistic higher spin wave equations, that is
Fierz-Pauli in the massive case and Fronsdal-Labastida in the massless one.
This analysis is done intrinsically, \emph{i.e.} working directly in $D$ dimension, without considering the dimensional reduction.
This is the strategy that we follow once we extend the model to (A)dS backgrounds.
Of course, keeping in mind the dimensional reduction simplifies a bit the derivation of the field equations in flat space.

Thus, we work in odd $D$ dimensional flat spacetime, with $D=2d-1$, $d\geq2$, and use the complex U($s$) covariant combinations for fermions as defined in section \ref{review}, that now read $(\psi_I^\mu,\bar\psi^{\mu I})$ and $(\theta_I,\bar\theta^I)$. As before, we represent $\psi$'s and $\theta$'s as multiplications by the corresponding Grassmann variable, and $\bar\psi$'s, $\bar\theta$'s as derivatives thereof. We indicate a generic state of the model by $\ket{\cal R}$, and we mostly work with the wave function
${\cal R}(x,\psi,\theta)=(\bra{x}\otimes\bra{\psi}\otimes\bra{\theta})\ket{{\cal R}}$, which has a finite Taylor expansion in $\psi_I^\mu$ and $\theta_I$, where $I=1,...,s$. Since the $\theta$ content will distinguish between different types of spacetime tensors, we find it convenient to isolate it explicitly writing the state as
\begin{equation}\label{theta expansion}
{\cal R}(x,\psi,\theta)=\sum_{n=0}^s\,\frac{1}{n!}\,R^{I_1...I_n}(x,\psi)\,\theta_{I_1}...\theta_{I_n}\;,
\end{equation}
with $R^{I_1...I_n}:= R^{[I_1...I_n]}$ being totally antisymmetric in the $I$ indices. Here and in what follows $[...]$ will always denote weighted antisymmetrization.
In the constraints the $(\psi,\de_\psi)$ and $(\theta, \de_\theta)$ parts play different roles: the first performs algebraic operations on the single tensors contained in $R^{I_1...I_n}$,
while the second mixes different tensor structures. For this reason we write the constraints
in split form as follows
\begin{equation}\label{split operators SO}
\begin{split}
{\cal J}_I^J &= \psi_I^\mu\frac{\de}{\de\psi_J^\mu}+\theta_I\frac{\de}{\de\theta_J}-d\,\delta_I^J=J_I^J+\theta_I\frac{\de}{\de\theta_J}-d\,\delta_I^J\;,\\[2mm]
{\cal K}_{IJ} &= \psi_I^\mu\psi_{J\,\mu}+\theta_I\theta_J=g_{IJ}+\theta_I\theta_J\;,\\[2mm]
\bar{{\cal K}}^{IJ} &= \frac{\de^2}{\de\psi_I^\mu\de\psi_{J\,\mu}}+\frac{\de^2}{\de\theta_I\de\theta_{J}}={\rm tr}^{IJ}+\frac{\de^2}{\de\theta_I\de\theta_{J}}\;,
\end{split}
\end{equation}
for the SO($2s$) algebra operators, and
\begin{equation}\label{split operators super}
\begin{split}
{\cal Q}_I &= \psi_I^\mu\,p_\mu+m\,\theta_I=Q_I+m\,\theta_I\;,\quad\bar{{\cal Q}}^I=p_\mu\,\frac{\de}{\de\psi_I^\mu}+m\,\frac{\de}{\de\theta_I}=\bar Q^I+m\,\frac{\de}{\de\theta_I}\;,\\[2mm]
{\cal H} &= \tfrac12\left(p^2+m^2\right)=H+\frac{m^2}{2}\;,
\end{split}
\end{equation}
for the supersymmetry part. We named $g_{IJ}:= \psi_I^\mu\psi_{J\,\mu}$ and ${\rm tr}^{IJ}:= \frac{\de^2}{\de\psi_I^\mu\de\psi_{J\,\mu}}$
the $D$-dimensional parts of ${\cal K}_{IJ}$ and $\bar{\cal K}^{IJ}$ in order to emphasize their algebraic meaning. The algebra of the $D$-dimensional operators $J_I^J$, $g_{IJ}$, ${\rm tr}^{IJ}$, $Q_I$, $\bar Q^I$ and $H$ is  the same as the massless algebra presented in section \ref{review} in $D+1$ dimensions,
up to the normal ordering constant that for simplicity we have not included in $J_I^J:= \psi_I^\mu\frac{\de}{\de\psi_J^\mu}$, but we give it here for completeness. The SO($N$) part reads
\begin{equation}\label{Dalgebra1}
\addtolength{\fboxsep}{5pt}
\begin{gathered}
[{ J}_I{}^J, { J}_K{}^L] =
\delta_K^J { J}_I{}^L -\delta_I^L {  J}_K{}^J
\\[1.5mm]
[{ J}_I{}^J, {g}_{KL}] =
 \delta_K^J {g}_{IL} + \delta_L^J {g}_{KI}
\\[1.5mm]
[{ J}_I{}^J, {\rm tr}^{KL}] =
 - \delta_I^K  {\rm tr}^{JL} - \delta_I^L {\rm tr}^{KJ}
\\[1.5mm]
[{g}_{IJ}, {\rm tr}^{KL}] =  4\delta_{[J}^{[k}\,J^{L]}_{I]}-(2d-1)\,\left(\delta_J^K\delta_I^L-\delta_I^K\delta_J^L\right)
\end{gathered}
\end{equation}
while the $R$-symmetry rotations are given by
\begin{equation}\label{Dalgebra2}
\addtolength{\fboxsep}{5pt}
\begin{gathered}
[{ J}_I{}^J,{ Q}_K] =
\delta_K^J { Q}_I
\\[1.5mm]
[{ J}_I{}^J,\bar { Q}^K] =
- \delta_I^K \bar { Q}^J
\\[1.5mm]
[ {\rm tr}^{IJ},{ Q}_K]=
 \delta_K^J \bar { Q}^I  - \delta_K^I \bar { Q}^J
\\[1.5mm]
[{ g}_{IJ},\bar { Q}^K] =
 \delta_J^K { Q}_I  - \delta_I^K { Q}_J  \ ,
\end{gathered}
\end{equation}
and the susy algebra is
\begin{equation}\label{Dalgebra3}
\{Q_I,\bar{Q}^J\} = 2 \delta_I^J { H} \ ,
\qquad \{{ Q}_I,{ Q}_J\}  = \{\bar { Q}^I,\bar { Q}^J\} =0 \ .
\end{equation}

To understand the tensor content of the various $R^{I_1...I_n}$ terms we recall that the diagonal ${\cal J}$'s are number operators
that count the numbers ${\rm N}_{\psi_I}$ of $\psi_I$'s and ${\rm N}_{\theta_I}$ of  $\theta_I$'s.
The constraints ${\cal J}_I^I{\cal R}=0$, where $I$ is fixed and not summed, amount then to
$$
\left({\rm N}_{\psi_I}+{\rm N}_{\theta_I}\right){\cal R}=d\,{\cal R}\;.
$$
This means that in ${\cal R}$
 we have $d$ antisymmetric indices in the $I$-th group whenever the $\theta_I$ is \emph{not} present, while we have $d-1$ antisymmetric indices when it is. From the decomposition \eqref{theta expansion}, it is thus clear that $R^{I_1...I_n}$ contains $n$ ``short'' columns\footnote{We refer to columns, using a Young tableau language, to denote blocks of antisymmetric indices.} of $(d-1)$ indices labeled by $I_1...I_n$, and the remaining $(s-n)$ ``long'' columns of $d$ indices, \emph{i.e.}
 \begin{equation}
R^{I_1...I_n}\sim\Yvcentermath1\bigotimes^{s-n}\,\young(\6,\5,\vdots,\7)\quad\bigotimes^n\,\young(\6,\vdots,\1)\;.
\end{equation}
This is covariantly stated by splitting in $\theta$ the ${\cal J}_I^J{\cal R}=0$ equation, that reads
\begin{equation}\label{J constraints}
\left(J_K^L-d\,\delta_K^L\right)\,R^{I_1...I_n}+n(-)^{n-1}R^{L[I_1..I_{n-1}}\delta^{I_n]}_K=0\;,
\end{equation}
and looking at its diagonal part.
The off-diagonal parts of these constraints plays two roles. First, they enforce GL($D$) irreducibility on each $R^{I_1...I_n}$ as a spacetime tensor. They further tell that, for given $n$, all the $\binom{s}{n}$ seemingly different $R^{I_1...I_n}$ actually represent the same spacetime tensor with the same Young tableau, and they only differ in the $\psi_I$ structure.
To see this it would be much easier to go back to the $(D+1)$-dimensional picture, but we can still analyze \eqref{J constraints} in a bit more detail.
For $K\neq L$ the operator $J_K^L$ removes a spacetime index from column $L$ and antisymmetrizes it within column $K$. Equation \eqref{J constraints} can be split in three cases:
\begin{itemize}
  \item $L\in\{I_1..I_n\}$: Removing an index from a short column,
 and placing it in any other column where it is antisymmetrized,  gives zero.
   \item $K,L\notin\{I_1..I_n\}$: Removing an index from a long column and antisymmetrizing it within a long column gives zero.
  \item $L\notin\{I_1..I_n\},\;K\in\{I_1..I_n\}$: Removing an index from a long column $L$ and antisymmetrizing it within a short column $K$ equates it to
   another $R^{I_1...I_n}$ tensor having a short $L$ column and long $K$ one.
\end{itemize}
The first two conditions amount to GL($D$) irreducibility, while the third says that the various $R^{I_1...I_n}$ at fixed $n$ differ only in naming which columns are the short ones, \emph{i.e.} they only differ in the $\psi_I$ species.

Summarizing the whole content of \eqref{J constraints} we have that, for given $n$, any $R^{I_1...I_n}$ is represented by the same spacetime tensor whose GL($D$) Young tableau is obtained from a rectangular $d\times s$ one by removing $n$ cells from the bottom row
\begin{center}\vspace{10mm}
\begin{picture}(0,60)(30,-35)
\put(0,-30){$\yng(6,6,6,3)$}
\put(0,-32){$\underbrace{\hspace*{42pt}}_{s-n}$}
\put(42,-18){$\underbrace{\hspace*{41pt}}_{n}$}
\put(-70,-7){$R^{I_1...I_n}\sim \scalebox{0.7}{$d$}\,\left\{\rule{0pt}{31pt}\right.$}
\end{picture}
\end{center}

The entire field content is then given by the $s+1$ tensors $\{R,R^I,R^{IJ},..,R^{I_1...I_s}\}$. Starting from the maximal rank one,
 $R$ with rectangular $d\times s$ Young tableau, one goes down by removing one by one the cells of the bottom row until $R^{I_1...I_s}$, with rectangular $(d-1)\times s$ diagram, is reached.
This final picture is clear having in mind the dimensional reduction of a tensor with a  rectangular Young tableau.

Having treated the tensor structure of the states of the physical Hilbert space, the other independent constraints, namely ${\cal Q}_I$ and $\bar{\cal K}^{IJ}$, give the dynamics for the system. The constraints on $R^{I_1...I_n}$ read
\be
\addtolength{\fboxsep}{5pt}
\boxed{
\begin{gathered}
Q_K\,R^{I_1...I_n}+m(-)^{sd+n+1}n\,\delta_K^{[I_1}R^{I_2...I_n]}=0
\\[1.5mm]
{\rm tr}^{KL}R^{I_1...I_n}-R^{KLI_1...I_n}=0 \;.
\end{gathered}
}
\label{Qtr constraints}
\ee
The first equation gives integrability conditions and, in the massive case, relates higher rank tensors to the lower rank ones via successive derivatives, while
the second equation enforces trace conditions that contain the truly dynamical equations.
The mass parameter in the integrability condition above gives different physical interpretations to the tensors $R^{I_1...I_n}$, depending whether it vanishes or not. For this reason we shall now treat separately the massive case and its massless limit.

\subsubsection{The massive case: Pauli-Fierz}
When the mass parameter is nonzero, we can invert the first equation in \eqref{Qtr constraints} to get higher rank curvatures in terms of lower ones
\begin{equation}\label{Qsolved}
R^{I_1...I_n}=(-)^{sd+n+1}\frac{1}{m(s-n)}\,Q_K\,R^{KI_1...I_n}\;.
\end{equation}
This can be iterated until they are all expressed in terms of
the last one $R^{I_1...I_s}$, that is the only independent field left, giving all the curvatures as
\vspace{5mm}
\begin{flushright}
\begin{picture}(0,60)(30,-35)\hspace{-35mm}
\put(0,-30){$\young(\hfill\hfill\hfill\hfill\hfill\hfill,\hfill\hfill\hfill\hfill\hfill\hfill,\hfill\hfill\hfill\hfill\hfill\hfill,\partial\partial\partial)$}
\put(0,-32){$\underbrace{\hspace*{42pt}}_{s-n}$}
\put(-1,24){$\overbrace{\hspace*{84pt}}^{s}$}
\put(-265,-.2){${\displaystyle R^{I_1...I_n}= \frac{(-)^{(s-n)(sd+1)}}{m^{s-n}(s-n)!}\,Q_{I_{n+1}}...Q_{I_s}\,R^{I_1...I_s}}\sim \scalebox{0.7}{$(d-1)$}\,\left\{\rule{0pt}{23pt}\right.$}
\end{picture}
\end{flushright}\vspace{-10mm}
\begin{equation}
\end{equation}\vspace{2mm}

Since every equation can be cast in terms of $R^{I_1...I_s}$ only, it is convenient to use the SU($s$) invariant symbol $\epsilon^{I_1...I_s}$ to dualize all the fields as
\begin{equation}\label{Us duality}
R^{I_1...I_n}=\epsilon^{I_1..I_nJ_1..J_{s-n}}\,\R_{J_1..J_{s-n}}\;.
\end{equation}
In particular we reserve a different name for the independent field: $\R=\phi$ (the one corresponding to $R^{I_1...I_s}$).
In this dual picture, the Young tableau for $\R_{I_1...I_n}$ is given by adding $n$ cells in a $d$-th row to the $(d-1)\times s$ box diagram of $\phi$
\vspace{5mm}
\begin{center}
\begin{picture}(0,60)(30,-35) 
\put(0,-30){$\young(\hfill\hfill\hfill\hfill\hfill\hfill,\hfill\hfill\hfill\hfill\hfill\hfill,\hfill\hfill\hfill\hfill\hfill\hfill,\hfill\hfill\hfill)$}
\put(-1,-32){$\underbrace{\hspace*{42pt}}_{n}$}
\put(-1,24){$\overbrace{\hspace*{84pt}}^{s}$}
\put(-90,-.5){${\displaystyle \R_{I_1...I_n}}\sim \scalebox{0.7}{$(d-1)$}\,\left\{\rule{0pt}{22.6pt}\right.$}
\end{picture}
\end{center}

For sake of completeness we give here all the relevant constraint equations for the fields in the dual basis
\begin{equation}\label{constraints dual basis}
\begin{split}
& \left(J_K^L-d\,\delta_K^L\right)\R_{I_1...I_n}+(n+1)\delta^L_{[K}\R_{I_1...I_n]}=0\;,\\[2mm]
& Q_K\,\R_{I_1...I_n}+m(-)^{sd}(n+1)\,\R_{KI_1...I_n}=0\;,\\[2mm]
& {\rm tr}^{KL}\R_{I_1...I_n}-\delta^K_{[I_1}\delta^L_{I_2}\R_{I_3...I_n]}=0\;.
\end{split}
\end{equation}
The first equation reproduces the field content just described.
The $Q$ equations can be solved iteratively to give
\begin{equation}\label{tildeR=QQQphi}
\R_{I_1...I_n}=\frac{(-)^{sd+1}}{m\,n}\,Q_{[I_1}\R_{I_2...I_n]}=\cdots=\frac{(-)^{n(sd+1)}}{m^nn!}\,Q_{I_1}...Q_{I_n}\phi\;,
\end{equation}
and the consistency condition $Q_K\R_{I_1...I_n}-Q_{[K}\R_{I_1...I_n]}=0$ is trivially satisfied due to the anticommuting nature of the $Q_I$'s\footnote{Note that on (A)dS spaces the $Q_I$'s do not anticommute,
and this will become a nontrivial consistency condition of the solution.}.

At this point the only fields needed are $\phi$ and the curvatures
\be
\R_I=\frac{(-)^{sd+1}}{m}\,Q_I\phi\;,\quad \R_{IJ}=\frac{1}{2m^2}\,Q_IQ_J\phi\;.
\ee
The relevant field equations come from the trace constraints in \eqref{constraints dual basis} for $n=0,1,2$ while all the higher order constraints will be derivatives of the field equations themselves. Explicitly, the relevant trace constraints are
\begin{equation}\label{relevant for PF}
\begin{split}
& {\rm tr}^{KL}\phi=0\;,\quad {\rm tr}^{KL}\R_I=0\;,\\[2mm]
& {\rm tr}^{KL}\R_{IJ}-\delta^K_{[I}\delta^L_{J]}\phi=0\;.
\end{split}
\end{equation}
The first equation tells that the field $\phi$ is completely traceless, while the second one reads ${\rm tr}^{KL}Q_I\phi=0$. Using the $[{\rm tr},Q]$ algebra in \eqref{Dalgebra2} and ${\rm tr}^{IJ}\phi=0$ one finds $\bar Q^I\phi=0$, \emph{i.e.} $\phi$ is divergence-free. At this point the last equation in \eqref{relevant for PF} becomes trivial for $\{KL\}\notin\{IJ\}$. The only nontrivial part sits in its contraction
\begin{equation}\label{KG}
{\rm tr}^{IJ}\R_{IJ}-\tfrac{s(s-1)}{2}\phi=0\;.
\end{equation}
By using the $[{\rm tr},Q]$ algebra as above and $\{Q_I,\bar Q^J\}=2\delta_I^J\,H$, as long as the previous trace and divergence constraints
are imposed, it simply becomes the massive Klein-Gordon equation
$$
(p^2+m^2)\phi=0\;.
$$

We have thus shown that the physical content of the model reduces to a single field $\phi$ described by a rectangular $(d-1)\times s$ Young tableau\vspace{10mm}
\begin{center}
\begin{picture}(0,60)(30,-35)
\put(0,10){$\yng(6,6,6)$}
\put(-1,8){$\underbrace{\hspace*{84pt}}_{s}$}
\put(-145,27){$\phi_{\mu_1..\mu_{d-1},...,\nu_1..\nu_{d-1}}\sim \scalebox{0.7}{$(d-1)$}\,\left\{\rule{0pt}{23pt}\right.$}
\end{picture}
\end{center} \vspace{-20mm}
\begin{equation}\label{explicit phi}
\end{equation}
obeying the Fierz-Pauli massive equations \cite{Fierz:1939ix}, that in our language read
\begin{equation}\label{FP}
\addtolength{\fboxsep}{5pt}
\boxed{
\begin{gathered}
{\rm tr}^{IJ}\phi=0\;,\quad\bar Q^I\phi=0\;,\quad\left(p^2+m^2\right)\phi=0\;.
\end{gathered}
}
\end{equation}
In the more explicit tensorial notation they take the form
\begin{equation}
\begin{split}
\phi^\mu{}_{\mu_2\dots\mu_{d-1},\dots,\mu\nu_2\dots\nu_{d-1}} &=0\\[1.5mm]
\partial^\mu \phi_{\mu\mu_2\dots\mu_{d-1},\dots,\nu_1\dots\nu_{d-1}} &=0\\[1.5mm]
\left( -\Box +m^2\right) \phi_{\mu_1\dots\mu_{d-1},\dots,\nu_1\dots\nu_{d-1}}  &=0 \;.
\end{split}
\end{equation}
Having analyzed the massive case, we can now turn to the somehow richer massless limit.

\subsubsection{The massless limit: Fronsdal-Labastida multiplets}

If we set the mass parameter to zero, the only constraint equations that change are the $Q$ ones. Since, as we will see, in this case one has $s+1$ different physical fields, there is no real advantage in using the dual basis, and we return to the original one, \emph{i.e.} $R^{I_1...I_n}$.
The irreducibility constraints ${\cal J}_I^J$ are exactly the same and as before they yield\vspace{15mm}
\begin{center}
\begin{picture}(0,60)(30,-35)
\put(0,10){$\yng(6,6,6,3)$}
\put(0,8){$\underbrace{\hspace*{42pt}}_{s-n}$}
\put(41.5,22){$\underbrace{\hspace*{41pt}}_{n}$}
\put(-70,33){$R^{I_1...I_n}\sim \scalebox{0.7}{$d$}\,\left\{\rule{0pt}{31pt}\right.$}
\end{picture}
\end{center}\vspace{-10mm}

The remaining equations now read
\begin{equation}\label{Qtr constraints massless}
\begin{split}
&Q_K\,R^{I_1...I_n}=0\;,\\[2mm]
&{\rm tr}^{KL}R^{I_1...I_n}-R^{KLI_1...I_n}=0\;.
\end{split}
\end{equation}
The $Q$ equations now tell us that each curvature separately obeys Bianchi integrability conditions, and indeed we shall integrate them in terms of $s+1$ different massless potentials. Part of the analysis is now strictly analogous to what we reviewed in section \ref{review}: we introduce two higher derivative operators
\begin{equation}\label{q,qIJ}
q=\frac{1}{s!}\,\epsilon^{I_1...I_s}Q_{I_1}...Q_{I_s}\;,\quad q^{IJ}=\frac{1}{(s-2)!}\,\epsilon^{IJI_3...I_s}Q_{I_3}...Q_{I_s}\;,
\end{equation}
and we use $q$ to solve the integrability constraints as
\begin{equation}\label{R=qphi}
R^{I_1...I_n}=q\,\varphi^{I_1...I_n}\;.
\end{equation}
To understand the tensor structure of $\varphi^{I_1...I_n}$, notice that $[J_I^J,q]=\delta_I^J\,q$. This means that at the level of Young tableaux $q$ attaches $s$ cells at the bottom of the $s$ columns of the diagram. Hence, the $\varphi^{I_1...I_n}$ Young tableau can be obtained from the tableau of $R^{I_1...I_n}$ by stripping off one cell from the bottom of each column. In general the resulting structure will be
\vspace{10mm}
\begin{center}
\begin{picture}(0,60)(30,-35)
\put(0,10){$\yng(6,6,3)$}
\put(0,8){$\underbrace{\hspace*{42pt}}_{s-n}$}
\put(42,22){$\underbrace{\hspace*{41pt}}_{n}$}
\put(-90,26.5){$\varphi^{I_1...I_n}\sim \scalebox{0.7}{$(d-1)$}\,\left\{\rule{0pt}{23pt}\right.$}
\end{picture}
\end{center}
\vspace{-10mm}
and the pictorial relation between curvature and potential is as follows\vspace{5mm}
\begin{center}
\begin{picture}(0,60)(30,-35)
\put(0,-30){$\young(\hfill\hfill\hfill\hfill\hfill\hfill,\hfill\hfill\hfill\hfill\hfill\hfill,\hfill\hfill\hfill\partial\partial\partial,\partial\partial\partial)$}
\put(0,-32){$\underbrace{\hspace*{42pt}}_{s-n}$}
\put(0,24){$\overbrace{\hspace*{84pt}}^{s}$}
\put(42,-18){$\underbrace{\hspace*{41pt}}_{n}$}
\put(-145,0){${\displaystyle R^{I_1...I_n}=q\,\varphi^{I_1...I_n}}\sim \scalebox{0.7}{$(d-1)$}\,\left\{\rule{0pt}{22.8pt}\right.$}
\end{picture}
\end{center} \vspace{2mm}
Something slightly different happens in $D=3$, that is $d=2$: the $(d-2)\times s$ box diagram is now empty, and one has symmetric tensors of spin ranging from zero to $s$, namely
$$
\varphi^{I_1...I_n}  \sim \
\quad
\underbrace{ \hspace{-1pt} \begin{array}{c}\\[-4.5mm]
\yng(3)
\end{array} }_{s-n}   \;,\quad D=3\;.
$$
Once the curvatures are written in terms of potentials, the field equations take the form
\begin{equation}\label{eom trqphi}
{\rm tr}^{KL}\,q\,\varphi^{I_1...I_n}-q\,\varphi^{KLI_1...I_n}=0\;,
\end{equation}
and they are higher derivative equations. Following the derivation sketched in section \ref{review}, we notice that
\begin{equation}\label{trq=qG}
\begin{split}
{\rm tr}^{KL}\,q &= q^{KL}\, G\;,\quad G=-2H+Q_I\bar Q^I+\tfrac12\,Q_IQ_J\,{\rm tr}^{IJ}\;\\[2mm]
q\,\varphi^{KLI_1...I_n}&= \frac12\,q^{KL}\,Q_IQ_J\,\varphi^{IJI_1...I_n}\;,
\end{split}
\end{equation}
where $G$ is the Fronsdal-Labastida operator. The equations \eqref{eom trqphi} can then be recast as
\begin{equation}\label{FL entangled}
q^{KL}\,\left(G\,\varphi^{I_1...I_n}-\tfrac12\,Q_IQ_J\,\varphi^{IJI_1...I_n}\right)=0\;.
\end{equation}
The expression inside the bracket mixes different potentials, but it is possible to decouple them recursively with a field redefinition:
\begin{equation}\label{field redef}
\varphi^{I_1...I_n}=\tilde\varphi^{I_1...I_n}+\sum_{j=1}^m\alpha_j^{(n)}\,g_{K_1L_1}...g_{K_jL_j}\,\tilde\varphi^{K_1L_1..K_jL_jI_1..I_n}\;.
\end{equation}
Namely, one starts from $\varphi^{I_1..I_s}=\tilde\varphi^{I_1..I_s}$ and $\varphi^{I_1..I_{s-1}}=\tilde\varphi^{I_1..I_{s-1}}$ and goes down until $\varphi^{I_1..I_n}$ with $n=s-2m$ or $n=s-1-2m$ in \eqref{field redef}. The $\alpha^{(n)}_j$ coefficients can be found by recursion and read
$$
\alpha^{(n)}_0:=1\;,\quad \alpha^{(n)}_j=\frac{\alpha^{(n+2)}_{j-1}}{4j\left(j+n-s+\tfrac12\right)}\;,
$$
so that
\be
\alpha^{(n)}_j=\frac{1}{4^j j! \prod_{l=1}^j (2j-l +n-s +\tfrac12)}\;.
\ee
Once we have $s+1$ decoupled equations
\be
q^{KL}\,G\,\tilde\varphi^{I_1...I_n}=0
\ee
we can drop the tildes and, since $q^{IJ}\sim Q^{s-2}$, we can locally parametrize its kernel as $Q^3\rho$, namely
\begin{equation}\label{FLcompensator}
G\,\varphi^{I_1...I_n}=Q_IQ_JQ_K\,\rho^{IJK|I_1..I_n}\;,
\end{equation}
that are nothing but Fronsdal-Labastida equations for the mixed symmetry tensors $\varphi^{I_1...I_n}$ with compensators.
We can see from \eqref{FLcompensator} that for each $\varphi^{I_1...I_n}$ there are different compensator structures: indeed their Young tableaux are obtained from the corresponding $\varphi$ diagrams by removing three cells in the $IJK$ columns. Since there is no symmetry relation between the two sets $IJK$ and $I_1..I_n$, one has different tensors for $\rho$ whether some $IJK$ coincide with some $I_k$ or not. In the next subsection we will provide some explicit examples of the various structures that can arise.

\subsubsection{Gauge invariance}

As expected in a theory of massless fields, the equations giving the curvatures in terms of the potentials admit a gauge symmetry that leaves the curvatures invariant. Since $q\,Q_I=Q_I\,q=0$ it is easy to see that $\delta R^{I_1...I_n}=0$ if we vary the gauge field as
\begin{equation}\label{gauge varphi}
\delta\varphi^{I_1...I_n}=Q_K\,\Lambda^{K|I_1...I_n}\;.
\end{equation}
As it was the case for the compensators, the Young diagram of the gauge parameter is obtained from the gauge field one by removing one cell in all possible ways. This produces different gauge parameters whether the index $K$ coincides or not with one of the $I_k$. The curvatures are then gauge invariant under \eqref{gauge varphi}, but $G\,\varphi^{I_1..I_n}$ is not. Nonetheless the compensated equations are gauge invariant if we give the compensators the following transformation rule
\begin{equation}\label{gauge compensator}
\delta\rho^{IJK|I_1...I_n}=\tfrac12\,{\rm tr}^{[IJ}\Lambda^{K]|I_1...I_n}\;.
\end{equation}
One can then partially gauge fix the theory setting the compensators to zero to recover the usual Fronsdal-Labastida equations
\begin{equation}\label{FL}
G\,\varphi^{I_1...I_n}=0\;,
\end{equation}
that are gauge invariant for traceless gauge parameters: ${\rm tr}^{[IJ}\Lambda^{K]|I_1...I_n}=0$.

Now we would like to give some explicit examples to clarify which tensor structures appear for various spins and dimensions. If we take for instance $s=4$ in $D=5$, \emph{i.e.} $d=3$ we have in the massive case the single field
$$
\phi\sim\Yvcentermath1\yng(4,4)\;,
$$
while in the massless limit one obtains the following multiplet of massless Fronsdal-Labastida gauge fields
\begin{equation}\nonumber
\begin{split}
&\varphi\sim\Yvcentermath1\yng(4,4)\;,\quad\varphi^I\sim\Yvcentermath1\yng(4,3)\;,\quad\varphi^{IJ}\sim\Yvcentermath1\yng(4,2)\;,\\[2mm]
&\varphi^{IJK}\sim\Yvcentermath1\yng(4,1)\;,\quad\varphi^{IJKL}\sim\Yvcentermath1\yng(4)\;.
\end{split}
\end{equation}
Taking for instance $\varphi^{IJ}$, its gauge invariance is parametrized by two different set of parameters $\Lambda^{K|IJ}$
$$
\Lambda^{K|IJ}\sim\Yvcentermath1\yng(4,1)\quad K\notin\{I,J\}\quad\text{or}\quad\Yvcentermath1\yng(3,2)\quad K\in\{I,J\}\;,
$$
while the compensators appearing in \eqref{FLcompensator} can be
$$
\rho^{KLM|IJ}\sim\Yvcentermath1\yng(3)\quad\text{or}\quad\Yvcentermath1\yng(2,1)
$$
whether one or two indices in $KLM$ coincide with $IJ$.

\section{Coupling to curved space: (A)dS manifolds}
In the previous section we have described a massive higher spinning particle in flat, odd spacetime dimensions,
together with its massless limit, by dimensionally reducing a massless model defined in a flat even-dimensional space.
It is known that the latter can be coupled to (A)dS spaces, and more generally to conformally flat spaces.
Thus, it is natural to investigate possible extensions of our model to curved spaces. We focus in particular to (A)dS backgrounds,
which are the ones that appear in the construction of the Vasiliev's interacting models.

We proceed as follows. We covariantize the constraints that define our model. Then,
considering a curved metric, we check if the algebra remains first class. If
that happens to be true, it means that the gauge symmetries defining the model
are not broken by the curvature, and the model is viable. Indeed we find  that a coupling to (A)dS is allowed.

In order to deform the quantum constraint algebra to include a $D$ dimensional curved target space,
we start from the SO($N$) generators, where the task is easy.
In this case we only need to use worldline fermions  with flat indices.
For this purpose we use the first part of the greek alphabet to indicate flat indices, \emph{i.e.} $\psi^\alpha_i$ with $\alpha= 0,1,...,D-1$,
(curved indices can then be obtained by using a vielbein $e_\mu{}^\alpha$ and its inverse
$e_\alpha{}^\mu$). We define the quantum SO($N$)  generators by
\begin{align}
{\cal J}_{ij} = \frac{i}{2}\big[ \psi^\alpha_{i}, \psi_{j\alpha} \big] +\frac{i}{2} \big[ \theta_{i}, \theta_{j} \big] :=  J_{ij} +  L_{ij}
\label{eq:J2}
\end{align}
which are seen to satisfy the SO($N$) algebra (indeed there is no change with respect to the calculation in flat space).
The ordering prescription used in (\ref{eq:J2})  is uniquely fixed by the  SO($N$) algebra.
Then, we covariantize the susy generators by replacing the linear momentum $p_\mu$
with the covariant momentum $\pi_\mu$, \emph{i.e.}
\begin{align}
{\cal Q}_i &= \psi^\alpha_i e_\alpha{}^\mu \pi_\mu  +\theta_i m:= Q_i +  \theta_i m\\
\pi_\mu&= p_\mu-\tfrac{1}{2}\omega_{\mu \alpha \beta} M^{\alpha\beta}
\end{align}
where $M^{\alpha \beta}=\frac{i}{2}\big[\psi^\alpha_i ,\psi^{\beta}_i\big]$ are the  Lorentz generators in the multispinor representation
and $\omega_{\mu \alpha \beta}$ the spin connection.
Since by definition the  $Q_i$'s
do not involve $\theta$'s, they are nothing but the  massless susy constraints appearing in ref.~\cite{Bastianelli:2008nm}. They satisfy the commutation rule
\begin{align} \label{QiQjM}
\big\{ Q_i, Q_j\big\} &= 2\delta_{ij}  H_0 +\frac{i}{2} \psi_i^\alpha \psi_j^{\alpha'} R_{\alpha\alpha'\beta\beta'} M^{\beta \beta'}
\end{align}
where $ H_0 = \frac12 \big( \pi^\alpha\pi_\alpha -i\omega^{\beta}{}_{\beta \alpha} \pi^\alpha\big)$
and $R_{\alpha\beta\gamma \delta}$ is the Riemann curvature tensor.
For the full susy constraints we thus have
\begin{align} \label{QiQjM'}
\big\{{\cal Q}_i, {\cal Q}_j\big\} &= 2\delta_{ij}\big(  H_0 +\tfrac12 m^2\big) +\frac{i}{2} \psi_i^\alpha \psi_j^{\alpha'} R_{\alpha\alpha'\beta\beta'} M^{\beta \beta'}
\end{align}
which imply that ${\cal Q}_i$,  ${\cal J}_{ij}$ and a suitably chosen hamiltonian constraint $\cal H$ cannot possibly form an algebra of first class constraints
for a generic background.
However restricting to maximally symmetric spaces
\be
R_{\alpha\beta\gamma\delta} = b\big(\eta_{\alpha\gamma}\eta_{\beta\delta}- \eta_{\alpha \delta}\eta_{\beta\gamma}\big)
\label{mss}
\ee
one finds
\begin{align} \label{QiQjMAdS}
\big\{{\cal Q}_i, {\cal Q}_j\big\} &= 2\delta_{ij}\big( \tilde H +\tfrac12 m^2\big) +\frac{b}{2} \Big( \delta_{ij} J_{kk'} J_{kk'} - J_{ik}  J_{jk} - J_{jk}  J_{ik} \Big)
\end{align}
where
\begin{align}
 \tilde H = H_0 -\frac{b}{4} J_{kk'}  J_{kk'} -bA(D)~.
\end{align}
The second term is an improvement term that is added and subtracted in~\eqref{QiQjMAdS} to achieve $[ \tilde H, Q_i]=[ \tilde H, J_{ij}]=0$, and $A(D) =(2-N)\frac{D}{8} -\frac{D^2}{8}$ is a quantum effect due to operatorial ordering.
The essential point to observe is that the right hand side of eq. (\ref{QiQjMAdS}) is
expressed with respect to the $J_{ij}$ operators that are not the constraints operators ${\cal J}_{ij}=J_{ij}+ L_{ij}$ that we need
to impose to get the massive HS equations of motion. So we ought to rewrite such relations in terms of the ${\cal J}_{ij}$ operators. Using the commutator rule $\{\theta_i,\theta_j\}=\delta_{ij}$, we find
\begin{align*}
& L_{ik} L_{jk} +  L_{jk}  L_{ik} =\frac{N-1}{2}\delta_{ij}\,,\quad
 L_{kk'} L_{kk'} = \frac{N(N-1)}{4}
\end{align*}
and
 \begin{align*}
& J_{ik} J_{jk} + J_{jk}  J_{ik} = {\cal J}_{ik} {\cal J}_{jk} +{\cal J}_{jk} {\cal J}_{ik} -2 L_{ik} {\cal J}_{jk} -2 L_{jk} {\cal J}_{ik}+ \frac{N-1}{2}\delta_{ij}\\
&J_{kk'}  J_{kk'}  = {\cal J}_{kk'} {\cal J}_{kk'}-2L_{kk'} {\cal J}_{kk'} +\frac{N(N-1)}{4} \;.
\end{align*}
Thus we can rewrite~\eqref{QiQjMAdS} as
\begin{align}
\big\{{\cal Q}_i, {\cal Q}_j\big\} &= 2\delta_{ij}\,{\cal H}\nonumber\\
& +\frac{b}{2} \Big[ \delta_{ij} ({\cal J}_{kk'}-2 L_{kk'}) {\cal J}_{kk'} -{\cal J}_{ik} {\cal J}_{jk} - {\cal J}_{jk} {\cal J}_{ik} +2 L_{ik}  {\cal J}_{jk} +2 L_{jk}  {\cal J}_{ik}\Big]
\end{align}
and recover a first class system with the hamiltonian constraint $\cal H$ defined by
\begin{equation}
\begin{split}
{\cal H} &= H_0+\tfrac12 m^2 -\frac{b}{4} \Big( {\cal J}_{kk'}-2 L_{kk'}\Big)  {\cal J}_{kk'} -b\Big(A(D) +\frac{N-1}{8}\Big)\\[2mm]
&=\tilde H+\tfrac12 m^2 +\frac{b}{16}(N-1)(N-2) \;.
\end{split}
\end{equation}
Indeed, since ${\cal H}$ and $\tilde H$ only differ by a constant term, we have
\begin{align}
[{\cal H}, {\cal Q}_i] = [{\cal H}, {\cal J}_{ij}]=0
\end{align}
and ${\cal H}$ is  a central element of the algebra (in fact, ${\cal H}$ separately commutes with $Q_i, \theta_i,  J_{ij},  L_{ij}$).

To summarize, we have seen that the constraints
${\cal J}_{ij}$, ${\cal Q}_i$,  and $\cal H$ identified above
form a first class system with nontrivial structure functions. The latter arise because of the curvature
of the (A)dS spaces, encoded in the parameter $b$ that is related to the scalar curvature by
$R=D(D-1)b$.  We conclude that the massive higher spinning particle can be defined on (A)dS spaces.

A classical action for the model can be written down immediately using the
classical limit of the above constraints, and the corresponding lagrangian read as
 \bea
L \eqa
p_\mu \dot x^\mu + \frac{i}{2} \psi_{i\alpha} \dot \psi_i^\alpha + \frac{i}{2} \theta_{i} \dot \theta_i
-e \underbrace{\frac12 (\pi_\mu \pi^\mu -\tfrac{b}{2}J_{ij}J_{ij}+ m^2)}_{{\cal H}^{cl}}
 \ccr
&&
- i \chi_i \underbrace{\Big ( e_\alpha{}^\mu \pi_\mu \psi^\alpha_i + m \theta_i\Big )}_{{\cal Q}_i^{cl}}
- \frac12 a_{ij} \underbrace{\Big (i \psi_i^\alpha \psi_{j\alpha} + i \theta_i \theta_j
\Big )}_{{\cal J}_{ij}^{cl}} \;.
\label{action3}
\eea

\subsection{Light cone analysis}\label{LCAdS}

The lagrangian (\ref{action3}) describes a massive HS particle in odd dimensional (A)dS spaces.
As stressed, the constraint algebra is first class, so that its gauging is consistent and the model is viable.
To check that the HS particle indeed carries nontrivial degrees of freedom, we perform a light cone analysis at the classical level.
It goes in a  way similar to the one presented earlier for flat space, and we highlight just the main points.

To proceed  we follow \cite{Metsaev:1999ui}.
For simplicity we set $b=-1$ in (\ref{mss})
and use the Poincar\`e parametrization of the AdS space with
$ds^2=(-(dx^0)^2 + (dx^1)^2 +...+ (dx^{D-3})^2 + dz^2 + (dx^{D-1})^2)/z^2$, where a special role is played by the coordinate
$x^{D-2}:= z$.
As before we set $x^\pm= (x^{D-1}\pm x^0)/\sqrt{2}$, so that
$ds^2= (2 dx^+ dx^- + dx^A dx^A+ dz^2)/z^2$ with $A =1,2,..D-3$,
 and consider $x^+$ as the light cone time. We also use an index $a=(A, D-2)$ that runs over $D-2$ values to include
 the one corresponding to the coordinate $x^{D-2}:=z$, so that one could  write the metric
 in the form $ds^2= (2 dx^+ dx^- + dx^a dx^a)/z^2$ as well.

 Now we make the gauge choice $x^+=\tau$. Correspondingly one may  solve the hamiltonian constraint
 ${\cal H}^{cl}$  by
 $p_+= -\frac{1}{2 z^2 \pi_-} ( z^2 \pi_a \pi_a+\frac{1}{2} J_{ij}^2 + m^2) +\frac12 \omega_{+\alpha\beta} M^{\alpha\beta}$.
In a similar way we set $\psi^+_i =0$
 by a gauge choice, and solve
 ${\cal Q}_i^{cl}$  by
 $ \psi^-_i = -\frac{1}{e_-{}^\mu \pi_\mu}(e_a{}^\mu \pi_\mu \psi^a_i +m \theta_i)$.
 This leaves a lagrangian of the form\footnote{On fermions the indices are to be considered as flat.}
\bea
L \eqa  p_- \dot x^- + p_a \dot x^a  +\frac{i}{2}  \psi_{ia} \dot \psi_i^a +\frac{i}{2}  \theta_{i} \dot \theta_i
+ p_+
- \frac12 a_{ij} \underbrace{\Big (i \psi_i^a \psi_{ja} +i \theta_i\theta_j \Big )}_{{\cal J}_{ij}}
\eea
where the remaining algebraic constraints  related to the SO($N$) charges are exactly  the same as the one present in the
flat space discussion. Their treatment proceeds in the same way, so that we conclude that
the number of degrees of freedom remains unchanged when passing from flat space to an AdS background.

\subsection{Covariant analysis: massive case}

We have shown that the deformed constraint superalgebra remains first class on (A)dS backgrounds. This ensures that, in the complex U(s) basis, the independent constraints ${\cal J}_I^J$, $\bar{\cal K}^{IJ}$, ${\cal Q}_I$ produce consistent and covariant dynamical equations.
In order to find them we split again the operators according to the $\theta$ content as done in section \ref{covariantsection}. The field content in (A)dS remains unchanged, at the level of $R^{I_1...I_n}$ tensors, since the SO($N$) generators and the corresponding constraint analysis are unmodified.
The dynamics is governed as in the flat case by the integrability and trace conditions as in \eqref{Qtr constraints}, the difference with respect to flat space being the covariant momenta inside the $Q_I$, that give non trivial anticommutators. Indeed, the algebra of $D$-dimensional operators reported in \eqref{Dalgebra1} and \eqref{Dalgebra2} remains unchanged in (A)dS, while \eqref{Dalgebra3} becomes
\begin{equation}\label{allQAdS}
\begin{split}
&\left\{Q_I,Q_J\right\}=b\,\left(g_{IK}\,J^K_J+g_{JK}\,J^K_I\right)\;,\quad \left\{\bar{Q}^I,\bar{Q}^J\right\}=-b\,\left({\rm tr}^{IK}\,J^J_K+{\rm tr}^{JK}\,J_K^I\right)\;,\\[2mm]
&\left\{Q_I,\bar Q^J\right\}=2\,\delta_I^J\,H_0-\frac{b}{2}\,\Big(J_I^KJ_K^J+J_K^JJ_I^K-2 g_{IK}{\rm tr}^{JK}+(4-4d-s)J_I^J-\delta_I^JJ^K_K\Big)\;,
\end{split}
\end{equation}
where we prefer to give the last relation in terms of the minimal $H_0$, since it is the operator represented by the minimal covariant laplacian $2H_0=-\nabla^2$.
In the massive case the analysis proceeds along the same steps described in section \ref{covariantsection}. The only independent field is $\R=\phi$, with rectangular $(d-1)\times s$ Young tableau, obeying
\begin{equation}\label{massive AdS}
\begin{split}
& {\rm tr}^{KL}\phi=0\;,\quad \frac{1}{m}\,{\rm tr}^{KL}Q_I\,\phi=0\;,\\[2mm]
& \frac{1}{2m^2}{\rm tr}^{KL}\,Q_I\,Q_J\,\phi-\delta^K_{[I}\delta^L_{J]}\phi=0\;.
\end{split}
\end{equation}
The remaining trace conditions are automatically satisfied, provided that the three above equations hold. This fact, along with the mutual consistency of the $Q_I$ integrability constraints, is highly nontrivial on (A)dS. The light cone analysis of the previous subsection guarantees that the propagating degrees of freedom are conserved with respect to the flat case, ensuring that the model is not empty in (A)dS.\\
The first equation tells us as before that $\phi$ is traceless\footnote{In this section on (A)dS backgrounds, we treat tensor fields with flat Lorentz indices. Every covariant equation can be rewritten using curved base indices by sending $\eta_{\alpha\beta}$ to $g_{\mu\nu}$.} and, since the $[{\rm tr},Q]$ algebra is unchanged, the second one turns again into a divergence constraint
\be
\bar Q^I\,\phi=0\quad\leftrightarrow\quad \nabla^\alpha\phi_{\alpha\alpha_2..\alpha_{d-1},...,\beta_1..\beta_{d-1}}=0\;.
\ee
The main difference with respect to the flat case appears in the third equation, due to the deformed $\{Q_I,\bar Q^J\}$ algebra, as can be seen from \eqref{allQAdS}.
In order to get the Klein-Gordon equation from \eqref{massive AdS}, let us manipulate the ${\rm tr}QQ$ term. Pushing the trace operator through the supercharges, and using ${\rm tr}^{IJ}\phi=\bar Q^I\phi=0$ we get
$$
{\rm tr}^{KL}\,Q_IQ_J\,\phi=2\delta_I^{[L}\left\{\bar Q^{K]},Q_J\right\}\phi\;.
$$
We use now the superalgebra \eqref{allQAdS}, along with the U($s$) constraint $J_I^J\phi=(d-1)\delta_I^J\phi$, to obtain
$$
2\delta_I^{[L}\left\{\bar Q^{K]},Q_J\right\}\phi=2\delta_I^{[L}\delta_J^{K]}\Big(2H_0+b(d-1)(d-1+s)\Big)\phi\;.
$$
Finally, inserting the above result in \eqref{massive AdS} yields
\be
\Big(-2H_0-b(d-1)(d-1+s)-m^2\Big)\phi=0\;,
\ee
\emph{i.e.} the covariant Klein-Gordon equation with the mass term shifted by a geometric contribution, thus completing the triplet of massive Fierz-Pauli conditions in (A)dS:
\begin{equation}\label{PF AdS}
\begin{split}
& \phi^\beta{}_{\alpha_2..\alpha_{d-1},...,\beta\,\beta_2..\beta_{d-1}}=0\;,\\
& \nabla^\alpha\phi_{\alpha\,\alpha_2..\alpha_{d-1},...,\beta_1..\beta_{d-1}}=0\;,\\
& \Big(\nabla^2-b(d-1)(d-1+s)-m^2\Big)\phi_{\alpha_1..\alpha_{d-1},...,\beta_1..\beta_{d-1}}=0~.
\end{split}
\end{equation}

We are now ready to analyze the massless limit, that is considerably more involved. In general we will not find Fronsdal-Labastida equations for mixed symmetry gauge fields, and we will limit ourselves to work out an explicit example.

\subsection{Massless limit: an example}
We give an interesting and non trivial example of what the model describes in the ``massless" limit $m=0$. In such a limit the dynamical equations in (A)dS reduce to
\begin{equation}\label{Qtr constraints massless AdS}
\begin{split}
&Q_K\,R^{I_1...I_n}=0\;,\\[2mm]
&{\rm tr}^{KL}R^{I_1...I_n}-R^{KLI_1...I_n}=0~.
\end{split}
\end{equation}
This form is the same as the flat case one, but now the integrability conditions $QR=0$ become  non trivial, because of the non-vanishing $\{Q,Q\}$ anti-commutator in \eqref{allQAdS}, that prevents the operator $q_0=\frac{1}{s!}\epsilon^{I_1...I_s}Q_{I_1}...Q_{I_s}$ to be annihilated by $Q_K$. In the massless even-dimensional (A)dS models of~\cite{Bastianelli:2008nm} the integrability conditions on HS curvature, described by rectangular GL($D$) Young tableaux, were solved.
However, unlike~\cite{Bastianelli:2008nm}, in the present work the (A)dS deformation seems to introduce complications for solving the integrability condition of some specific curvature tensors, those described by ``pistol-shaped" Young tableaux. This is presumably related to the fact that in AdS the  pistol-shaped HS potentials have less gauge symmetries than the corresponding flat space ones\footnote{We thank Per Sundell for this observation.}
\cite{Brink:2000ag, Boulanger:2008up, Boulanger:2008kw}.
Nonetheless, a light cone analysis applied to the pistol-shaped HS curvature equations, indicates that the corresponding multiplet carries the same number of degrees of freedom in all maximally-symmetric  spaces; we will further comment on this point later on.

In the following we discuss the spin-two case  in AdS${}_3$, where again we refer to the spin as the number of columns in the Young tableaux representing the higher spin curvatures. Solving the algebraic U($2$) constraints $J_{I}^{~J}$, the field content of the model is given by the following curvatures
 \begin{center}
 \begin{tabular}{rcl}
 $R$& $\sim $& $\Yvcentermath1\yng(2,2)$\\[7mm]
 $R^I$& $\sim $& $\Yvcentermath1\yng(2,1)$\\[7mm]
 $R^{IJ}$& $\sim $& $\Yvcentermath1\yng(2)$
 \end{tabular}
 \end{center}
 We start analyzing the differential constraints on the first and last curvature since we expect the equations of motion to mix them as in the flat case (being rectangularly shaped we expect no particular difficulties to arise). The differential Bianchi identities $Q_K R=0=Q_KR^{IJ}$ can indeed be solved by introducing a symmetric rank-two gauge potential $\varphi$ and a scalar $\varphi^{IJ}$ as follows
 \be
 R=q_{(2)}\varphi\qquad\qquad
 R^{KL}=q_{(2)}\varphi^{KL}
 \ee
 with
 \be
 q_{(2)}=\frac 1 2 \epsilon^{IJ}\left(Q_IQ_J-b\,g_{IJ}\right)~.
 \ee
 We now use the algebra presented in the previous section to push the trace operator ${\rm tr}$ through $q_{(2)}$. Firstly let us introduce the AdS generalization of the Fronsdal-Labastida operator $G_{(2)}^{AdS}$ defined by
 \begin{align}\nonumber
& G^{AdS}_{(2)}:=\tfrac 1 2\epsilon_{KL}{\rm tr}^{KL}q_{(2)}\\[2mm]
 &=\Big(-2H_0+Q_I\bar{Q}^I+\frac 1 2 Q_{I}Q_J{\rm tr}^{IJ}-bg_{IJ}{\rm tr}^{IJ}+\frac b 2 J_K^{~L}J_{L}^{~K}-b(d+1)J_{K}^{~K}+b(2d-1)\Big)
 \end{align}
 The equations of motion for the above potentials thus are the trace constraints of~\eqref{Qtr constraints massless AdS}. For the scalar potential  $\varphi^{IJ}:=\epsilon^{IJ}\phi$ we simply get
 \be
 {\rm tr}^{IJ}q_{(2)}\varphi^{KL} =\epsilon^{IJ}G^{AdS}_{(2)}\varphi^{KL}=0\quad \Rightarrow\quad (\nabla^2-3b)\phi=0
 \ee
 where on the right hand side we explicitly evaluated the action of the SO($4$) constraints on the scalar potential. Therefore the equations of motions for $R^{IJ}$ leave one propagating degree of freedom (DoF), that of a scalar field.

 For the rank-two tensor, $\varphi=\frac 1 2 \phi_{\mu\nu} \epsilon^{IJ}\psi_I^\mu\psi_J^\nu$, the trace condition amounts to
 \begin{equation}\label{spin2eom}
 {\rm tr}^{KL}q_{(2)}\varphi=q_{(2)}\varphi^{KL}
 \end{equation}
and, similarly to the flat space counterpart, we aim to combine  $\varphi$ and $\varphi^{KL}$ in a single field. To achieve this task we introduce
 \be
 \widetilde{\varphi}:=\varphi-\frac 1 2g_{IJ}\varphi^{IJ}
 \ee
 and use the operatorial relation
 $$
 [G^{AdS}_{(2)},g]=\epsilon^{IJ}Q_IQ_J(3-2d+J_L^{~L})+(2d-3)g-bgJ_L^{~L}
 $$
 with $g=\epsilon^{IJ}g_{IJ}$ to make $G^{AdS}_{(2)}$ act on $\varphi^{IJ}$, so that, when we put $\varphi^{IJ}$ on shell, equation~\eqref{spin2eom} reduces to
 \be
 G^{AdS}_{(2)}\widetilde{\varphi}=0
 \ee
 which is the Fronsdal equation in (A)dS for a massless spin-two potential, \emph{i.e.}
\begin{equation}
\nabla^2\phi_{\mu\nu}-\nabla_\mu\nabla^\rho\phi_{\rho\nu}-\nabla_\nu\nabla^\rho\phi_{\rho\mu}+\nabla_\mu\nabla_\nu\phi^\rho_\rho+2b\left(g_{\mu\nu}\,\phi^\rho_\rho
-\phi_{\mu\nu}\right)=0
\end{equation}
that in three dimensions carries no DoF's (we dropped the tildes for simplicity).

Finally we now try to integrate $R^I$ and impose the trace constraint to extract the last equation of motion. After considering a natural class of ansatze, we could not find a nontrivial solution to $Q_K R^I=0$, and it seems that $R^I$ cannot be integrated in terms of a potential. Nonetheless, a direct light cone check shows that the constraints $Q_J R^I = {\rm tr}^{JK} R^I=0$ leave one propagating DoF, as it happens in flat space. As we mentioned above, the fact that AdS gauge potentials with non rectangular Young tableaux carry more DoF's than corresponding flat space potentials might be the origin of this difficulty in integrating the curvatures $R^{I_1...I_n}$ for $n\neq0,s$\footnote{Remember that $R$ and $R^{I_1...I_s}$ have rectangular Young tableaux and can thus be integrated.}. Indeed, the light cone analysis of sec. \ref{LCAdS} shows that the whole sum over fields $R^{I_1...I_n}$ has the same degrees of freedom in both flat and AdS spaces. One can thus conclude that integrating $R^{I_1...I_n}$ in AdS, if possible, would not give rise to the same gauge potentials as those introduced in flat space, namely $\varphi^{I_1...I_n}$.
We summarize the results for such spin-two case in the following table
 $$
 \begin{tabular}{c|ccc}
 Curvature&$R\sim \Yvcentermath1 \yng(2,2)$&$R^I\sim\Yvcentermath1 \yng(2,1)$&$R^{IJ}\sim\Yvcentermath1 \yng(2)$\\[7mm]
 &$\scalebox{1.5}{$\Downarrow$}$&$\scalebox{1.5}{$\Downarrow$}$&$\scalebox{1.5}{$\Downarrow$}$\\[5mm]Potential&$\widetilde{\varphi}\sim\Yvcentermath1 \yng(2)$&not integrable&$\varphi^{IJ}\sim {\rm scalar}$\\[5mm]EoM&
 $~~G^{AdS}_{(2)}\widetilde{\varphi}~~$=0& $~~Q_J R^I = {\rm tr}^{JK} R^I=0~~$  &$~~G^{AdS}_{(2)}\varphi^{IJ}=0~~$\\[5mm] DoF& 0 & 1 & 1
 \end{tabular}
 $$
The total number of DoF's is two, just like those of a massive spin-two in $D=3$.

\section{Conclusions}

We have constructed a relativistic action for a massive particle with higher spin by dimensionally reducing a massless model.
The massless model used, the spinning particle with local SO($N$) extended susy on the worldline,
propagates degrees of freedom only in a spacetime of even dimension, so that the emerging  model lives in a spacetime
of odd dimensions. We have covariantized it to introduce a coupling to (A)dS spaces,
and shown that the physical degrees of freedom propagating at the quantum level  satisfy the Fierz-Pauli equations
extended to (A)dS spaces. The massless limit of the model contains the same number of degrees of freedom.
Its covariant description, arising for the quantum Dirac constraints  related to the gauge symmetries
of the particle action, has a geometric interpretation in terms of curvatures, but
a reformulation in terms of gauge potentials is generically more
complex that the one arising in flat space, and we have just presented the simple example of $s=2$ in $D=3$.
We have not produced a general analysis of the massless case to uncover if and how the (A)dS geometrical equations
 are related to massless gauge potentials.
Indeed, it is also conceivable that some of the degrees of freedom  could be realized in the form of
``partially massless states" discovered in~\cite{Deser:1983mm, Deser:2001us}, and further analyzed in~\cite{Hallowell:2005np}, where the authors derived the generating function
of HS (A)dS actions  for both massive and partially massless fields by applying the log radial reduction technique
\cite{Biswas:2002nk}
to a one-dimension higher massless HS theory, an idea that was further developed within the tractor and BRST set up in~\cite{Grigoriev:2011gp}.
It could be interesting to see if and how partially massless states can be described in first quantization.

One could proceed further in the first quantized description of our model by considering a closed worldline, so to analyze the corresponding
one loop effective action.
Path integrals on curved spaces need a regularization \cite{Bastianelli:2006rx},
but they can be used successfully in worldline approaches to QFT problems
 \cite{Bastianelli:2002fv}. We expect  that for the
present model  the counterterms identified in \cite{Bastianelli:2011cc} are enough for  carrying out the perturbative evaluation
of the one loop effective action.
One might also expect that an exact evaluation be possible, as similar result have been
found on AdS spaces for higher spin fields \cite{Camporesi:1993mz, Camporesi:1994ga}.

Another direction where to extend  the present work is to consider dimensional reduction in more than one dimension.
This may allow to find  worldline actions that describe propagation of several multiplets of massive and massless HS excitations in flat and  AdS spaces.
Of course, it would be extremely interesting to see how to make the various HS particles self-interact in a first quantized picture,
though this is a  sensibly harder problem.

\acknowledgments{The authors are grateful to P. Sundell for discussions. They acknowledge the partial support of UCMEXUS-CONACYT grant CN-12-564. RB thanks the Universidad Andr\'es Bello for hospitality.  EL acknowledges partial support of SNF Grant No. 200020-149150/1.}


\end{document}